\documentclass[a4paper,11pt]{article}
\usepackage[usenames]{color}

\usepackage{amsfonts,amssymb}
\usepackage{theorem}
\usepackage[dvips]{graphicx}
\usepackage{picins}

\def\G{\Gamma}

\def\cuv{\frac{1}{(4\pi)^2} \frac{1}{D-4}}

\begin{document}

%

\rightline{IFUM-900-FT }

\Large
\bf
\centerline{One-loop Self-energy and Counterterms}
\centerline{in a Massive Yang-Mills  Theory based}
\centerline{on the Nonlinearly Realized Gauge Group}
\normalsize \rm

\large
\rm
\vskip 1.3 truecm
\centerline{D.~Bettinelli
\footnote{e-mail: {\tt daniele.bettinelli@mi.infn.it}}, 
R.~Ferrari\footnote{e-mail: {\tt ruggero.ferrari@mi.infn.it}}, 
A.~Quadri\footnote{e-mail: {\tt andrea.quadri@mi.infn.it}}}

\normalsize
\medskip
\begin{center}
Dip. di Fisica, Universit\`a degli Studi di Milano\\
and INFN, Sez. di Milano\\
via Celoria 16, I-20133 Milano, Italy
\end{center}

\vskip 0.7  truecm
\normalsize
\bf
\centerline{Abstract}
\rm
\begin{quotation}
\normalsize
In this paper we evaluate the self-energy of the vector mesons 
at one loop in our recently proposed subtraction scheme 
for massive nonlinearly realized SU(2) Yang-Mills theory.
We check the fulfillment of physical unitarity.
The resulting self-mass can be compared with the
value obtained in the massive Yang-Mills theory based on the
Higgs mechanism, consisting in extra terms due to the presence
of the Higgs boson (tadpoles included).  
Moreover we evaluate all the one-loop counterterms necessary for the
next order calculations. By construction they satisfy all the equations
of the model (Slavnov-Taylor, local functional equation and Landau
gauge equation). They are sufficient to make all the one-loop
amplitudes finite through the hierarchy encoded in the local
functional equation. 
\end{quotation}

\newpage

\section{Introduction}
\label{sec:intr}
We have recently proposed a subtraction procedure \cite{Bettinelli:2007tq} 
of the divergences in the  $SU(2)$  Y-M theory \cite{Yang:1954ek} 
with a mass term \cite{Veltman:1968ki} - \cite{Ferrari:2004yt}  
based on a nonlinearly realized gauge group. 
This theory has no Higgs boson in the perturbative approach.

The proposed subtraction scheme is based on the following strategy.
i) A local functional equation is derived  encoding a hierarchy among the 1-PI 
 Green functions. According to this hierarchy all the amplitudes involving at
 least one unphysical Goldstone boson are fixed by the local functional 
 equation once one knows the amplitudes independent of the Goldstone bosons
 (ancestor amplitudes). ii) It is shown that only a finite number of 
 divergent ancestor amplitudes exists at every loop order 
 (weak power-counting).
 iii) The subtraction of the divergences is based on dimensional 
   regularization.  In particular the local functional equation 
 indicates that only the poles in $D - 4$ should be removed in the 
 properly normalized amplitudes.   
  
Thus the algorithm does not modify the number of the independent 
parameters of the zero-loop effective action. Hence, although the 
original langrangian is not renormalizable, we construct 
order by order in $\hbar$ a
consistent theory which depends on three parameters: the coupling constant
$g$, the mass $M$ and the   mass scale $\Lambda$
for the radiative corrections.
The tree-level vertex functional compatible with the symmetry properties 
of the theory (Slavnov-Taylor, local functional equation  
and Landau gauge equation) and the weak power-counting is unique.
This strategy is unconventional and departs
from the standard renormalization procedure.
\par
The proof of consistency (in the iterative subtraction) has been given
in a series of papers \cite{Bettinelli:2007tq}-\cite{Bettinelli:2007kc}. 
In particular the Slavnov-Taylor identity \cite{ST} is mantained
after the counterterms are introduced. The same is valid for the local
functional equation (LFE) derived from the transformation properties under
local left multiplication as well as for the Landau gauge equation.
\par
Physical unitarity is guaranteed to follow from the Slavnov-Taylor
identity \cite{physunit}, \cite{Ferrari:2004pd}.
Locality of the counterterms follows from the above mentioned local functional
equation. The construction of the counterterms is 
based on two important properties of this
equation: hierarchy and weak power-counting, which allow a full control
of the amplitudes involving the auxiliary scalar fields (descendant) in 
terms of the amplitudes with no auxiliary fields (ancestor). 
\par
In this work we provide as an example the evaluation
of the self-energy of the vector meson in $D$ dimensions
by using the Landau gauge. 
This explicit calculation is necessary for the following reasons:
i) to show how the proposed subtraction procedure works;
ii) to check that the Landau gauge (because of its unphysical 
pole at zero mass) does not pose any problem for physical unitarity
as it is required for the proof given in Ref.~\cite{Ferrari:2004pd};
iii) to provide the quantitative difference between the theories with
the linear (with Higgs boson) and the nonlinear representation (no Higgs boson)
of the gauge group.

The result shows how physical unitarity
is recovered on-shell. A comparison with the theory where the gauge group 
is linearly realized (Higgs mechanism \cite{ssb}) is very interesting. It shows
that our approach yields a consistent identification of the Higgs part.
This can be done on the physically relevant part: the self-mass of the
vector meson.  The discussion of this item necessitates the comparison
of our calculation with previuos works \cite{Marciano:1980pb} 
usually employing a 't Hooft gauge \cite{'tHooft:1971fh} .
Since the tadpoles (vacuum expectation value of the Higgs field) are gauge
dependent, the comparison can be made only for the self-mass.
The identification of the Higgs contribution, up to the mass scale
$\Lambda$, is possible because our approach does not allow the introduction
of free parameters for each local invariant solution of the defining
equations. These solutions for the one-loop case are listed in Appendix
\ref{app.C}.
\par
The self-energy of the vector boson can be evaluated at the two-loop
level. This calculation necessitates of the local one-loop counterterms.
In this paper we evaluate all the counterterms necessary for any two-loop
calculation. This amounts to find the coefficients of the pole
parts in $D-4$ for all the ancestor amplitudes, i.e. for all external
legs $A_\mu, V_\mu, K_0,\Theta_\mu, c, \bar c$ and $A_\mu^*,c^*,\phi^*,
\phi_0^*$.  Counterterms with Goldstone boson external legs are obtained from
those involving only ancestor variables. They satisfy the
linearized ST identity, the linearized LFE and the Landau gauge equation.
These constraints imply non trivial relations
among the ancestor amplitudes in the sector
spanned by the external sources
and the ghost field.
Finally the counterterms are described by a suitable basis of invariant
local solutions of the same equations.
Their coefficients are evaluated from the divergent part
of the ancestor amplitudes. The latter are collected in Appendix~\ref{app.div}.

\section{Effective action at the tree level and counterterms}

\label{sec:FR}
The Feynman rules are implicitly given by
the vertex functional at the tree level
\begin{eqnarray}&&
\Gamma^{(0)}
=\frac{\Lambda^{(D-4)}}{g^2} \int d^Dx \, \Biggl\{
- \frac{1}{4}  G_{a\mu\nu} G^{\mu\nu}_a +
\frac{M^2}{2} (A_{a\mu} - F_{a\mu})^2  
\nonumber \\&&
+  B_a (D^\mu[V](A_\mu - V_\mu))_a
- \bar c_a (D^\mu[V] D_\mu[A] c)_a +\Theta_{a}^\mu~(D_\mu[A]\bar{c})_a 
\nonumber \\& & 
+ \Big ( A^*_{a\mu} sA^\mu_a + 
\phi_0^* s \phi_0 + 
\phi_a^* s \phi_a + c_a^* s c_a
+ K_0 \phi_0 \Big )
\Biggr\}  \, .
\label{FR.1}
\end{eqnarray}
where, beside the conventional notations, $B_a$ is the Lagrange
multiplier for the Landau gauge,  $V_{a\mu},
\Theta_{a\mu}, K_0 $ are
the external sources necessary for the LFE and $ A^*_{a\mu},
\phi_0^*,\phi_a^*,c_a^*$ are the anti-fields for the BRST-transforms
$sA^\mu_a,s \phi_0, s \phi_a, s c_a$. The mass scale $\Lambda$
enters as a common factor in order to simplify the subtraction
procedure. The nonlinearity of the representation
of the gauge group $SU(2)_{\rm LEFT~LOCAL}\otimes SU(2)_{\rm
  RIGHT~GLOBAL}$
comes from the constraint on $\phi_0$:
\begin{eqnarray}&&
F_{a\mu}\frac{\tau_a}{2}=F_\mu = i \Omega \partial_\mu\Omega ^\dagger \qquad
\Omega_{ij} = \frac{1}{v}(\phi_0+i \tau_a\phi_a)_{ij}\in SU(2)
\nonumber\\&&
\phi_0= \sqrt{v^2-\vec\phi^2}.
\label{FR.1.0}
\end{eqnarray}
The complete set of Feynman rules includes
the counterterms: 
\begin{eqnarray}
\widehat\G\equiv \Gamma^{(0)}+ \frac{\Lambda^{(D-4)}}{g^2} 
\sum_{j\ge 1}\int d^Dx {\cal M}^{(j)}.
\label{FR.2}
\end{eqnarray}
The  counterterms ${\cal M}^{(j)}$ are given 
by the pole parts in $D-4$ of the normalized vertex functional
(the normalization is tightly conditioned by the particular
form of the effective action in eq.(\ref{FR.1})):
\begin{eqnarray}
\int d^Dx {\cal M}^{(j)}=-
\frac{g^2}{\Lambda^{(D-4)}}\sum_{k=0}^{j-1}\Gamma^{(j,k)}\biggl|_
{\rm POLE~PARTS}
\label{FR.3}
\end{eqnarray}
where $\Gamma^{(j,k)}$ denotes the vertex functional where the
total power of $\hbar$ of the inserted counterterms is $k$.
The subtraction procedure is consistent if the counterterms are
local and if the relevant equations are preserved: Slavnov-Taylor,
LFE and Landau gauge equation. We have given the formal proofs
that the subtraction proposed in eq. (\ref{FR.2}) works
for the Feynman rules in eq. (\ref{FR.1}) \cite{Bettinelli:2007tq}.
In the present paper we provide an explicit one-loop
calculation. The result will be compared with the result
of the linear theory (Higgs mechanism). In our final formula
it is evident how physical unitarity is realized and how the
parameter $v$ of eq. (\ref{FR.1.0}) disappers from the
final result since it is not a physical parameter 
\cite{Bettinelli:2007tq}.
\par
\begin{figure}
\begin{center}
\includegraphics[width=1.0\textwidth]{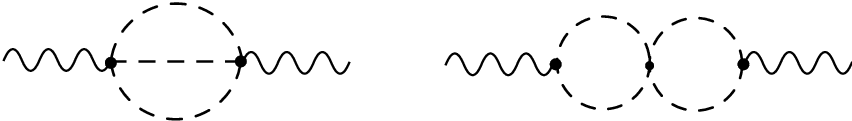}
\end{center}
\caption{Two-loop graphs where the nonlinearity appears.
Dashed lines are $\vec\phi$.}
\label{fig.1}
\end{figure}

We provide also all the one-loop counterterms for the ancestor
amplitudes (those with no $\phi$-external-legs). A complete
two-loop calculation is expected to be a straightforward task,
without any obstruction in the subtraction procedure,
since the structure of the divergences is, for most graphs,
that of the linear theory. There are few exceptions as those
depicted in Fig. \ref{fig.1}, which however have been already
consistently dealt with in the nonlinear sigma model \cite{Ferrari:2005ii}.
%
\section{Self-energy for the nonlinear massive YM}
\label{sec:selfnl}
\begin{figure}
\begin{center}
\includegraphics[width=1.0\textwidth]{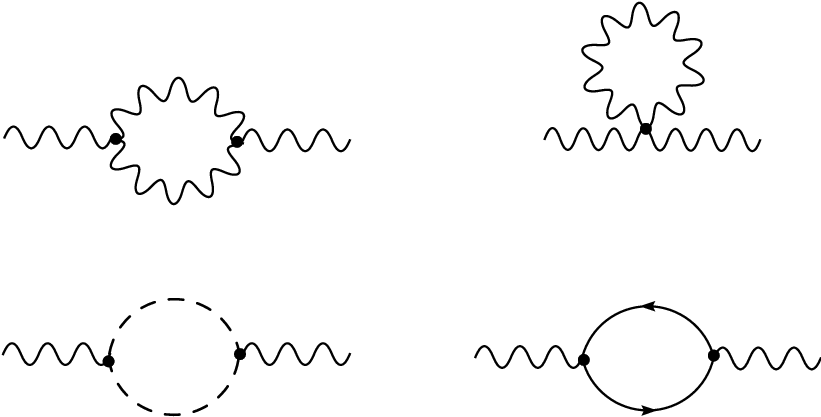}
\end{center}
\caption{Graphs of self-energy for the nonlinear theory.
Arrows are for FP ghosts.}
\label{fig.2}
\end{figure}

We give here the complete result of the one-loop
calculation in $D$ dimensions, without any subtractions.
The graphs are shown in Fig. \ref{fig.2}.
The transverse part is
\begin{eqnarray}&&
\Sigma_T(p^2) 
\!\!=-\frac{i}{D-1}
\nonumber\\&&
\!\!\Biggl\{
H_1(M^2)
\biggl[-2(D-1)^2+
2D-\frac{5}{2}+(2D-3)\frac{p^2}{M^2}
-\frac{p^2}{2M^2}\Big(1-\frac{M^2}{p^2}\Big)^2\biggr]
\nonumber\\&&
\!\!
+H_2(M^2,M^2)\biggl[(-2D+3)\frac{p^4}{M^2}+(7D-10)p^2
+4(D-1)M^2
\nonumber\\&& 
~~~~~~~-p^2\Big(\frac{p^2}{2M^2}-1\Big)^2 \biggr]
\nonumber\\&& \!\!
+ H_2(M^2,0) \left[\left( M^2(2D-3)+\frac{p^2}{2}\right)
\Big(\frac{p^2}{M^2}-1\Big)^2 
+\frac{p^2}{2}\Big(1-\frac{M^2}{p^2}\Big)^2\right]
\nonumber\\&&\!\!
+\frac{p^2}{4} \biggl(1-\frac{p^4}{M^4}\biggr)H_{2}(0,0)\Biggr\}
\label{selfnl.1}
\end{eqnarray} 
where
\begin{eqnarray}&&
H_{1}(m^2)=\int_{\cal M} 
\frac{d^Dq}{(2\pi)^D}\frac{1}{(q^2-m^2)}
\nonumber\\&&
H_{2}(m^2_1,m^2_2)=\int_{\cal M}  
\frac{d^Dq}{(2\pi)^D}\frac{1}{(q^2-m^2_1)[(p-q)^2-m^2_2]}\, .
\label{selfnl.2}
\end{eqnarray} 
The longitudinal part is
\begin{eqnarray}&&
\Sigma_L(p^2)
  =  -i H_1(M^2) \Big(  
\frac{3}{2}-\frac{p^2}{2M^2}
\Big)
\nonumber\\
&&
+i\frac{p^2}{2}
\Big(1-\frac{M^2}{p^2}\Big)^2\biggl[H_2(M^2,0)-\frac{1}{M^2}H_1(M^2)\biggr] 
-i\frac{p^2}{2}H_2(0,0) .
\label{selfnl.3}
\end{eqnarray} 

It is worth to notice some points:
\begin{enumerate}
\item $ \Sigma_T(0)=\Sigma_L(0)$ is verified for generic $D$. 
By this property the pole at $p^2=0$ in the 1PI two-point function is avoided.
This condition is very important in order to prove  physical unitarity
in the Landau gauge \cite{Ferrari:2004pd}.
\item For $p^2=M^2$,  $ \Sigma_T$ contains only $H_2(M^2,M^2)$ which is the
only Feynman integral with a physical discontinuity across
the real positive $p^2$ axis.
\item 
As a check on  $ \Sigma_L(p^2)$ the relevant Slavnov-Taylor 
identity is
explicitly evaluated in Appendix \ref{app:st}.
\end{enumerate}
The self-mass around $D=4$ can be evaluated according to the
prescription of  eq. (\ref{FR.3}). One gets
\begin{eqnarray}
g^2\Sigma_T (M^2)|_{D\sim 4}
=g^2\frac{M^2}{(4\pi)^2}\Biggl\{-\frac{23}{4}
C_\Lambda
+\frac{2}{3}
-\frac{33}{4}\int_0^1 \! dx\, \ln P(1,x)\Biggr\}
\label{selfnl.4}
\end{eqnarray} 
with 
\begin{eqnarray}
C_\Lambda\equiv\frac{2}{D-4}+\gamma
- \ln 4\pi+\ln\left(\frac{M^2}{\Lambda^2}\right)
\label{selfnl.5}
\end{eqnarray} 
and 
\begin{eqnarray}
P(r,x)\equiv x^2-rx+r.
\label{selfnl.6}
\end{eqnarray} 
%
\section{Self-energy in  the linear theory}
\label{sec:lin}
\begin{figure}
\begin{center}
\includegraphics[width=1.0\textwidth]{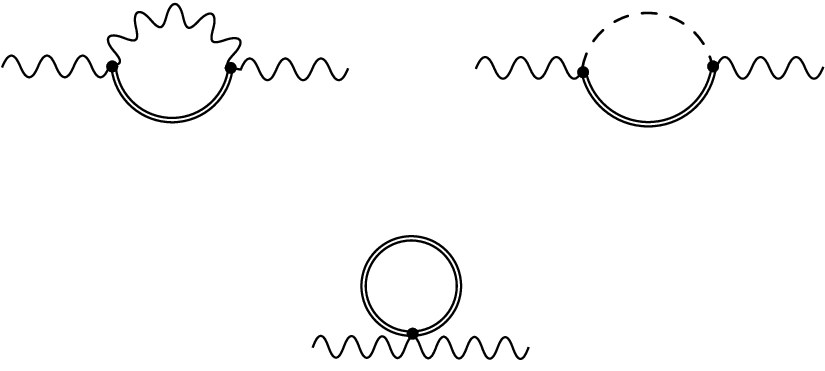}
\end{center}
\caption{Graphs of self-energy for the linear theory (involving a Higgs line)}
\label{fig.3}
\end{figure}

At one loop it is straightforward to evaluate 
the contribution of the Higgs sector. By this
we mean the contribution of the graphs in Fig. \ref{fig.3}.
Our approach fixes the separation of the Higgs from the
non-Higgs contribution once $\Lambda$ is given. This is at variance
with other approaches where the Higgs part is removed by hand.
In these methods the arbitrariness introduced at one loop 
is due to the presence of free parameters associated to the local
solutions of the ST identity and of the linearized LFE,
once the logs of $M_H$ are removed by hand.
This problem has been discussed
thoroughly in Refs. \cite{Bettinelli:2007zn}, \cite{Bettinelli:2007kc} 
for the nonlinear sigma model.
The non-decoupling
effects in the large Higgs mass limit have been
  studied at length in the literature (see e.g.
Refs.~\cite{Herrero:1994iu} for the Standard Model and \cite{Dittmaier:1995cr} 
for the SU(2) case).

\par
The Higgs contribution to the self-energy is evaluated in the
Landau gauge by using  the same form for the effective action
of eq. (\ref{FR.1})  without the constraint in eq. (\ref{FR.1.0}).
The mass term becomes ($\phi_0=h+v$)
\begin{eqnarray}&&
\frac{M^2}{2} (A_{a\mu} - F_{a\mu})^2 
= M^2 Tr\left( A_\mu -i\Omega\partial_\mu\Omega^\dagger \right)^2
\nonumber \\&&
= M^2Tr\Bigr\{\biggl[
\Omega^\dagger A_\mu - i \partial_\mu \Omega^\dagger
\biggr]\biggl[
 A_\mu \Omega+ i \partial_\mu \Omega 
\biggr]\Biggr\}
\nonumber\\&&
=\frac{4M^2}{v^2}\Biggl\{\frac{1}{2}\partial_\mu h \partial^\mu h +
\frac{1}{2}\partial_\mu \phi_a \partial^\mu\phi_a
+\frac{1}{8}A^2(h^2 +2vh +v^2 + \vec \phi^2)
\nonumber\\&&
+\frac{1}{2}A_a^\mu \biggl[\partial_\mu h \phi_a
- (h+v) \partial_\mu \phi_a +\epsilon_{abc}\phi_b \partial_\mu\phi_c 
\biggr]\Biggr\}
\label{lin.1}
\end{eqnarray}
and the potential is added
\begin{eqnarray}&&
-\frac{\lambda^2}{4}\left(h^2+\vec\phi^2 +2vh\right)^2
\nonumber\\&&
=-\lambda^2v^2h^2 - \frac{\lambda^2}{4}\left(
h^4+\vec\phi^4 + 2h^2\vec\phi^2  +4vh^3
+4vh\vec\phi^2\right).
\label{lin.2}
\end{eqnarray}
The Higgs mass is
\begin{eqnarray}
M_H^2=\frac{\lambda^2v^4}{2M^2}.
\label{lin.3}
\end{eqnarray}
By using these Feynman rules the contribution of the graphs
in Fig. \ref{fig.3} is evaluated. The contribution of the Higgs 
 to the transverse part of the two-point function is
\begin{eqnarray}&&
\!\!\!\!\!\!\!\!\!\!\!\!\!\!
\Sigma^{\rm HIGGS}_{T}(p^2) 
=
-\frac{i}{4}\frac{1}{(D-1)}
\nonumber\\&&
\Biggl\{
H_1(M^2_H)\Big(\frac{M_H^2}{p^2}-\frac{M^2}{p^2}+2-D\Big)
- H_{2}(M^2,M^2_H)\biggl[4(D-2)M^2
\nonumber\\&&
+\frac{(p^2 +M^2-M_H^2)^2}{p^2}\biggr]
+H_1(M^2)\Big(\frac{M^2}{p^2}-\frac{M_H^2}{p^2}+1\Big)
\Biggr\}
\label{ms.10.1}
\end{eqnarray} 

\begin{figure}
\begin{center}
\includegraphics[width=0.7\textwidth]{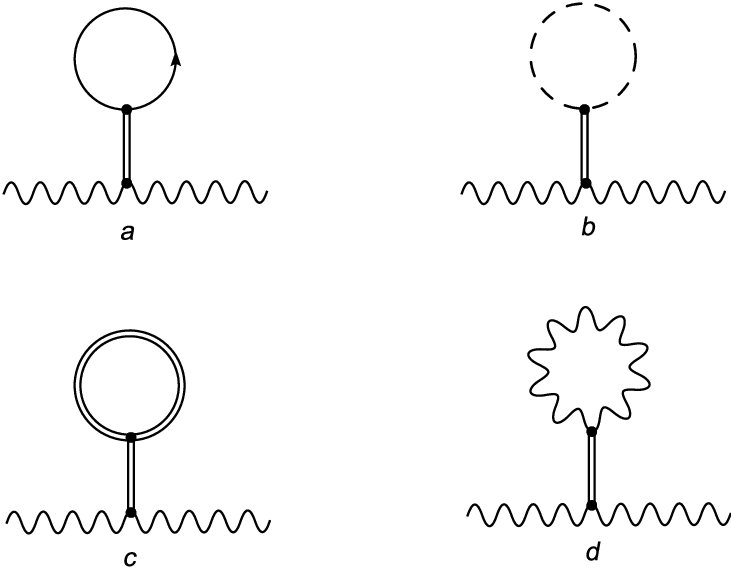}
\end{center}
\caption{Tadpoles originated from the nonzero vev of $\phi_0$. a) and b)
(tadpoles of the F-P ghosts and of the Goldstone boson) are zero for
massless Goldstone bosons.}
\label{fig.4}
\end{figure}

The contribution of the Higgs sector to the longitudinal
part of the two-point function is
\begin{eqnarray}&&\!\!\!\!\!\!\!\!\!\!\!\!\!\!\!\!\!\!\!\!\!
\Sigma^{\rm HIGGS}_{L}(p^2) 
=-\frac{i}{4}\Biggl[ 
\frac{M^2-M_H^2}{p^2} H_1(M^2_H)
- \frac{p^2 +M^2-M_H^2}{p^2} H_1(M^2)
\nonumber\\&&\!\!\!\!\!\!\!\!\!\!\!\!\!\!\!\!\!\!\!\!\!\!\!\!
+\biggl[\frac{(p^2 +M^2-M_H^2)^2}{p^2}- 4 M^2\biggr]H_{2}(M^2,M^2_H)
+(2M_H^2-p^2) H_{2}(M^2_H,0)\Biggr] \, .
\label{ms.10.2}
\end{eqnarray} 
For later discussion let us remind that eqs. (\ref{ms.10.1})
and (\ref{ms.10.2}) are the contribution of the Higgs
sector (in the linear theory) to the self-energy
of the vector meson in the Landau gauge.
The graphs shown in Fig. \ref{fig.4} have to be included as a contribution
coming from the Higgs sector. 
In fact we want to compare the predictions
in the linear and nonlinear realization of the gauge group. Thus
no finite subtraction is performed and the parameters entering
in the self-mass ($g$ and $M$) are given by the zero-order values.
In the Landau gauge both graphs a and b in Fig. \ref{fig.4} are zero.
Then we have
\begin{eqnarray}
&& \Sigma_{T}^{\rm TADPOLES}(p^2)  =  \Sigma_{L}^{\rm TADPOLES}(p^2)
\nonumber \\&&
~~~~~ = -\frac{3i}{4}H_1(M^2_H)
   -\frac{3i}{2}\frac{M^2}{M^2_H}(D-1)H_1(M^2).
\label{dt.5}
\end{eqnarray}
%

\section{Self-mass in the Nonlinear versus Linear  }
\label{sec:comp}
The results of the Sections \ref{sec:selfnl} and \ref{sec:lin}
allow a comparison of the self-mass in the two cases.
 The subtraction procedure of the
poles in $D-4$ must be the same. However this is not enough. In the
linear case a finite renormalization is always posssible and in particular
it is possible to drop the tadpole contributions, since they can
be eventually accounted for by some mass counterterms. A comparison
between the theories based on the linear and the nonlinear representation 
of the gauge group necessitates that the parameters $g$ and $M$ enter
as zero-order values and not as dummy variables. In fact the counterterms
can be even gauge-dependent if they are introduced in order to
balance the dropping of the tadpoles (see next section). 
\par
Now we take the case $p^2=M^2$ of $\Sigma_T$ and then consider the
Laurent expansion around $D=4$. We have in the nonlinear case
\begin{eqnarray}&&
\delta M^2_{\rm NONLINEAR}
=\frac{g^2M^2}{(4\pi)^2}\biggl\{-\frac{23}{4}\textit{C}_\Lambda
+\frac{2}{3}
-\frac{33}{4}\int_0^1 \! dx\, \ln P(1,x)\biggr\}\,.
\label{comp.1}
\end{eqnarray}
The linear theory, based
on the Higgs mechanism, adds to the above term the following
quantity (tadpoles of Fig. \ref{fig.4} are included) 
\begin{eqnarray}&&
\delta M^2_{\rm LINEAR}=
\delta M^2_{\rm NONLINEAR}+
\frac{g^2}{4}\frac{M^2}{(4\pi)^2}
\biggl\{\left(\frac{10}{3}-3r-\frac{18}{r} \right)\textit{C}_\Lambda
+\frac{10}{9}
\nonumber\\&&
\!\!\!\!\!\!\!\!\!\!\!\!\!\!
+\frac{6}{r}+\frac{7}{3}r+\frac{r^2}{3}
-\Big ( 2r+\frac{r^2}{3} \Big ) \ln r
+\Bigl(4-\frac{4}{3}r+\frac{r^2}{3}\Bigr )\int_0^1 dx \ln P(r,x)
\biggr\} 
\label{comp.2}
\end{eqnarray}

where 
\begin{eqnarray}
r=M^{-2}M_H^2.
\label{comp.3}
\end{eqnarray}
%
\section{$\langle 0|\phi_0|0\rangle$ is gauge dependent}
\label{sec:gi}
The comparison of our calculation, given in eqs. (\ref{comp.1}),
and (\ref{comp.2}), with results present in 
the literature needs some consideration about gauge
invariance of the vacuum. Thus we use  the 't Hooft gauge-fixing 
for the linear theory
\begin{eqnarray}&&
{\cal L}_{'t~Hooft}
=
\frac{B_a^2}{2\alpha}+B_a\left(\partial A_a+\frac{2M^2}{v\alpha}\phi_a
\right)
\nonumber\\&&
-\bar c_a\left(\partial^\mu D[A]_{ab\mu} +\frac{M^2}{v\alpha}
(\phi_0\delta_{ab}-\epsilon_{abc}\phi_c)
\right)c_b.
\label{gi.1}
\end{eqnarray}
Hereafter we list the amplitudes for the tadpoles in Fig. \ref{fig.4}
\begin{eqnarray}&&
\Sigma_{aa'\mu\nu}^{\rm TADPOLE ~ FP}(p)
= \frac{3i}{2}\frac{1}{\alpha}\frac{M^2}{M^2_H}g_{\mu\nu}\delta_{aa'}
H_1\Big(\frac{M^2}{\alpha}\Big)
\nonumber\\&&
\sim  \frac{3}{2}\frac{1}{\alpha}\frac{M^2}{M^2_H}g_{\mu\nu}\delta_{aa'}
\frac{M^2}{\alpha (4\pi)^2}\left(
\frac{2}{D-4}-1+\gamma-\ln 4\pi+\ln \frac{M^2}{\alpha \Lambda^2}\right).
\label{gi.2}
\end{eqnarray}
\begin{eqnarray}&&
\Sigma_{aa'\mu\nu}^{\rm TADPOLE ~GOLDSTONE}(p)
=-\frac{3i}{4}g_{\mu\nu}\delta_{aa'}H_1\Big(\frac{M^2}{\alpha}\Big)
\nonumber\\&&
\sim - \frac{3}{4}g_{\mu\nu}\delta_{aa'}
\frac{M^2}{\alpha (4\pi)^2}\left(
\frac{2}{D-4}-1+\gamma-\ln 4\pi+\ln \frac{M^2}{\alpha\Lambda^2}\right).
\label{gi.3}
\end{eqnarray}
\begin{eqnarray}&&
\Sigma_{aa'\mu\nu}^{\rm TADPOLE~ HIGGS}(p)
= -\frac{3i}{4}g_{\mu\nu}\delta_{aa'}H_1(M^2_H)
\nonumber\\&&
\sim -\frac{3}{4}g_{\mu\nu}\delta_{aa'}\frac{M_H^2}{(4\pi)^2}\left(
\frac{2}{D-4}-1 +\gamma-\ln 4\pi+\ln \frac{M_H^2}{\Lambda^2}\right)
\label{gi.4}
\end{eqnarray}
\begin{eqnarray}&&
\Sigma_{aa'\mu\nu}^{\rm TADPOLE ~ GAUGE}(p)
\nonumber\\&&
= -\frac{3i}{2}\frac{M^2}{M^2_H}g_{\mu\nu}\delta_{aa'}
\left((D-1)H_1(M^2)+ \frac{1}{\alpha}H_1\Big(\frac{M^2}{\alpha}\Big)\right)
\nonumber\\&&
\sim - \frac{3}{2}\frac{M^2}{M^2_H}g_{\mu\nu}\delta_{aa'}
\Biggl\{(D-1)
\frac{M^2}{ (4\pi)^2}\left(
\frac{2}{D-4}-1+\gamma-\ln 4\pi+\ln \frac{M^2}{\Lambda^2}\right)
\nonumber\\&&
+\frac{1}{\alpha}\frac{M^2}{\alpha (4\pi)^2}\left(
\frac{2}{D-4}-1+\gamma-\ln 4\pi+\ln \frac{M^2}{\alpha \Lambda^2}\right)
\Biggr\}.
\label{gi.5}
\end{eqnarray}

It is amazing that
\begin{eqnarray}&&
\Sigma_{aa'\mu\nu}^{\rm TADPOLE ~ GAUGE}(p)
+\Sigma_{aa'\mu\nu}^{\rm TADPOLE ~ FP}(p)
\nonumber\\&&
= -\frac{3i}{2}\frac{M^2}{M^2_H}g_{\mu\nu}\delta_{aa'}
(D-1)H_1(M^2)
\label{gi.6}
\end{eqnarray}

is gauge-independent. Moreover
\begin{eqnarray}&&
\langle 0|\phi_0|0\rangle = v\Biggl\{1  
-\frac{g^2}{2M^2}\frac{1}{\Lambda^{(D-4)}}\frac{3i}{4} 
\biggl[H_1(M^2_H)+2\frac{M^2}{M^2_H}
(D-1)H_1(M^2)
\nonumber\\&&
~~~~~~~~~~~~~~~~~ + 
H_1\Big(\frac{M^2}{\alpha}\Big)
\biggr]\Biggr\}
\label{gi.7}
\end{eqnarray}
i.e. the vev of $\phi_0$ is gauge-dependent through the mass of
the Goldstone boson $\frac{M^2}{\alpha}$.
\par
The above discussion shows that tadpoles have to be considered
in the evaluation of the self-mass if one wants a gauge-invariant
result. In the linear theory it is not compelling to introduce
the tadpoles, since one can always perform a finite renormalization
in order to restore gauge invariance. However, with this choice,
one is not allowed to use the physical parameters for
the zero-order-value entries of $g$, $M$ and $M_H$.
\par
The comparison of our results with the expression given
by Marciano and Sirlin in Appendix A Ref. \cite{{Marciano:1980pb}} must 
take into accounts these facts. Their result for the gauge group
$SU(2)$
\footnote{This equation has been obtained by using the identity
$$
 \int_0^1 dx\, P(r,x) \ln P(r,x)
=\frac{1}{3}\Biggl(
- \frac{2}{3} + r-\frac{r^2}{2}+\frac{r^2}{2}\ln r
-\Big(\frac{r^2}{2}-2r \Big) \int_0^1\! dx \,\ln P(r,x)
\Biggr)
$$}
\begin{eqnarray}&&
A(M^2)_{AA}^{\rm MARCIANO~SIRLIN}
= \frac{g^2M^2}{16\pi^2}\Biggl\{- \frac{25}{6}C_\Lambda
+\frac{7}{36}
\nonumber\\&&
-\frac{1}{6} \Bigl ( r-\frac{r^2}{2} \Bigr )
 - \frac{r^2}{12}\ln r
+\frac{r}{4} \ln r
\nonumber\\&&
-\frac{33}{4}\int_0^1\! dx\, \ln P(1,x)
+ \Biggl( \frac{r^2}{12}-\frac{r}{3}
+1 \Biggr)  \int_0^1\! dx\, \ln P(r,x)\Biggr\}
\label{marciano.15}
\end{eqnarray} 
must be complemented by the contribution of the tadpole b in Fig. \ref{fig.4}
(at $\alpha=1$) in order to get a gauge-invariant result
\begin{eqnarray}&&
A(M^2)_{AA}^{\rm MARCIANO~SIRLIN}+\Sigma^{\rm TADPOLE ~ GOLDSTONE}
\nonumber\\&&
= \frac{g^2M^2}{16\pi^2}\Biggl\{- \frac{59}{12}C_\Lambda
+\frac{34}{36}
-\frac{1}{6}
\Bigl ( r-\frac{r^2}{2} \Bigr )
- \frac{r^2}{12}\ln r
+\frac{r}{4} \ln r
\nonumber\\&&
-\frac{33}{4}\int_0^1\! dx\, \ln P(1,x)
+\Biggl( \frac{r^2}{12}-\frac{r}{3}
+1
\Biggr)
\int_0^1\! dx\, \ln P(r,x)
\Biggr\}
\label{marciano.15p}
\end{eqnarray} 
This agrees with our results in eq. (\ref{comp.2}) if we add to the
expression in eq. ({\ref{marciano.15p}}) the contributions of
the gauge-, Higgs- and Faddeev-Popov-tadpoles as reported in eqs. (\ref{gi.4})
and (\ref{gi.6}).

\section{One-loop Counterterms}\label{sec:oc}

Two-loop calculations require the knowledge of the
full set of one-loop counterterms. The counterterms must obey
the ST identity, the local functional equation and
the Landau gauge equation
\cite{Bettinelli:2007tq}. According to the hierarchy property,
only the counterterms involving ancestor variables have to be
computed in order to implement the iterative subtraction
of the divergences.
The full list of the relevant invariant solutions
is reported in Appendix \ref{app.C} specialized to the case
where the descendant fields are neglected (Goldstone boson fields).
Counterterms involving descendant field external legs are obtainable
by using the full expression of the invariant solutions given
in Ref. \cite{Bettinelli:2007tq}: the compact expressions,
written in terms of bleached fields, must be projected on the
relevant monomials.

The coefficients of the invariants are determined
by computing the divergent part 
of the relevant ancestor amplitudes
after the proper normalization given by eq.(\ref{FR.3}).
The divergences of the ancestor amplitudes
are collected in Appendix~\ref{app.div}.

One finds
\begin{eqnarray}
\widehat \G^{(1)} &= & \frac{\Lambda^{(D-4)}}{(4\pi)^2} \frac{1}{D-4}
\Big [  \frac{17}{2} ({\cal I}_1 - {\cal I}_2)  -\frac{67}{6}
{\cal I}_3 + \frac{11}{4} {\cal I}_4 - \frac{5}{2} {\cal I}_5
+ 3 M^2 {\cal I}_6  \nonumber \\
& & ~~~~~~~~~~~~~~~~~ 
-6 {\cal I}_7 
+ \frac{3v^2}{128 M^4} {\cal I}_8 
-  \frac{v}{8 M^2} {\cal I}_9 \Big ]
\, .
\label{oc.3}
\end{eqnarray}
${\cal I}_{10}$ and ${\cal I}_{11}$  do not enter
into the parameterization of the one-loop counterterms. This 
is a peculiar property of the Landau gauge. 
\par
It is clear from eq. (\ref{oc.3}) that the one-loop counterterms
for the pure gauge sector cannot be casted in the form
\begin{eqnarray}
{\cal I}_1-{\cal I}_2-2{\cal I}_3+{\cal I}_4-{\cal I}_5
=\frac{1}{4}\int d^Dx \, G_a^{\mu\nu} G_{a\mu\nu},
\label{oc.4}
\end{eqnarray}
as noted already in the early works on the divergences of
the pure massive Y-M theory \cite{Shizuya:1976qf}, 
\cite{Kafiev:1982hx}. Our approach allows to overcome
this difficulty by managing the divergences with another
set of tools based both on BRST transformations and invariance
of the path integral measure under local
left multiplication.

Despite the fact that they are divergent by power-counting,
one-loop 1-PI amplitudes involving more than one $V$-leg are finite.
This result can be established from eq.(\ref{oc.3}) by noticing
that the dependence of $\widehat \G^{(1)}$ on $V$ is only linear  
(via the invariant ${\cal I}_7$).

\section{Conclusions}

In this paper we have provided the $D$-dimensional self-energy of
the vector meson in the SU(2) gauge group
in the nonlinearly realized perturbative formulation 
recently proposed in \cite{Bettinelli:2007tq}. 
We have discussed how physical unitarity is recovered
on-shell and presented a comparison with the linear
theory. Such a comparison is possible since
the subtraction scheme of \cite{Bettinelli:2007tq} allows to separate
the Higgs part of the self-mass. 
This is a consequence of the fact that in our
approach no free parameters can be introduced
for each local invariant solution of the defining equations,
as listed in Appendix \ref{app.C} for the one-loop case.
We have also given the full set of one-loop counterterms 
which are required for any  two-loop computation.
The counteterms have been parameterized in terms
of invariant solutions of the ST identity, the LFE and
the landau gauge equation. Their coefficients are obtained
from the evaluation of the divergent part of the ancestor amplitudes
(no Goldstone fields).

\section{Acknowledgments}

We are indebted with Glenn Barnich and Stefan Dittmaier for very 
stimulating discussions.
We acknowledge a partial financial support by MIUR.

\appendix
\section{One-loop Invariants}\label{app.C}

We list here the eleven invariants compatible
with the symmetry requirements and the WPC 
at the one loop level. By the Landau gauge equation the dependence
of $\G^{(1)}$ on $\bar c_a$  happens in the combination
\begin{eqnarray}
\widehat A^*_{a\mu} = A^*_{a\mu} +  (D_\mu[V]\bar c)_a \, . 
\end{eqnarray}
We neglect the descendant fields.
\begin{eqnarray}
&& {\cal I}_1 = \frac{1}{2} 
\int d^Dx \, \partial_\mu A_{a\nu} \partial^\mu A^\nu_a \, ,
\nonumber \\
&& {\cal I}_2 = \frac{1}{2} \int d^Dx \, (\partial A_a)^2 \, , \nonumber \\
&& {\cal I}_3 = 
- \frac{1}{2} \int d^Dx \, \epsilon_{abc} \partial_\mu A_{a\nu} A^\mu_b A^\nu_c
\, ,
\nonumber \\
&& {\cal I}_4 = \frac{1}{4} \int d^Dx \, (A^2)^2  \, , 
\nonumber \\
&& {\cal I}_5 = \frac{1}{4} \int d^Dx \, (A_{a\mu} A_b^\mu) 
(A_{a\nu} A_b^\nu) \, , \nonumber \\
&& {\cal I}_6 = \frac{1}{2} 
\int d^Dx \, A^2 \,  , \nonumber \\
&& {\cal I}_7 = 
\frac{1}{2}
\int d^Dx \,  
V_a^\mu \Big (  
 D^\rho G_{\rho \mu}[A]  + M^2 A_\mu \Big )_a 
- \frac{1}{2} 
\int d^Dx \, {\widehat A}^*_{a\mu} \Theta_a^\mu
\nonumber \\
&& ~~~~~~~~~~~~~ + \frac{1}{2} \int d^Dx \,  {\widehat A}^*_{a\mu}
(D^\mu[V] c)_a 
\,  ,
\nonumber \\
&& 
{\cal I}_{8} = 
\int d^Dx \, (2 K_0 - c_a \phi_a^*)^2 \, , 
\nonumber \\
&&
{\cal I}_{9} 
=   \int d^Dx \, \Big ( \frac{1}{2} c_a \phi_a^* A^2
- K_0 A^2 \Big ) \, , 
\nonumber \\
&& {\cal I}_{10} = 
\int d^Dx \, \Big ( 
\frac{1}{2}  (D^\mu[A] {\widehat A}^*_\mu)_a c_a - \frac{1}{4}
\phi_a^* c_a - \frac{1}{2} c_a^* \epsilon_{abc} c_b c_c 
\Big )  \, ,
 \nonumber \\
&& 
{\cal I}_{11} = 
\int d^Dx \, \Big ( c_a \phi_a^* - 2 K_0 \Big ) \, .
\label{anc.inv}
\end{eqnarray}
We remind, once again, that ${\cal I}_{1}-{\cal I}_{11}$ are not
solutions of the ST identity, local functional equation, and 
Landau gauge equation. Instead, they are the projection on the ancestor
variables of the solutions given in Ref. \cite{Bettinelli:2007tq}.

\section{One-loop divergences of the ancestor amplitudes}\label{app.div}

\renewcommand{\labelenumi}{Graph (\alph{enumi})}

In this Appendix we give the one-loop divergent
parts of the ancestor amplitudes. The resulting counterterms are for 
the theory where the gauge group is represented nonlinearly and 
in the Landau gauge. The Feynman rules are encoded in 
eq.(\ref{FR.1}). 
From that action we can read immediately the free propagators  
(~the factor $\frac{g^2}{\Lambda^{(D-4)}}$ is always left understood~)
%
%
\begin{itemize}
\item[]
\parpic{\includegraphics[width=2cm]{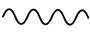}} 
$~~\Delta_{A_{a\mu}A_{b\nu}} = \frac{-i }{p^2 -  M^2} \Big ( g_{\mu\nu} 
- \frac{p_\mu p_\nu}{p^2} \Big ) \delta_{ab}$
$~$ \\
\item[]
\parpic{\includegraphics[width=2cm]{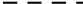}}
$~~\Delta_{\phi_a \phi_b} 
= \frac{i}{4}\frac{ v^2}{ M^2} 
\frac{1}{p^2}\, \delta_{ab} $
$~$ \\
\item[]
\parpic{\includegraphics[width=2cm]{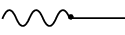}}
$~~\Delta_{B_a A_{b\mu}} = \frac{p_\mu}{p^2}\, \delta_{ab}$
$~$ \\ 
\item[]
\parpic{\includegraphics[width=2cm]{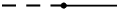}}
$~~\Delta_{B_a \phi_b} = - i \frac{v}{2 p^2} \, \delta_{ab}$
$~$ \\
\item[]
\parpic{\includegraphics[width=2cm]{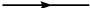}}
$~~\Delta_{c_a \bar c_b} 
= \frac{i}{p^2}\, \delta_{ab}$
$~$ \\
\item[]
\parpic{\includegraphics[width=2cm]{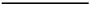}}
$~~\Delta_{B_a B_b} =0$
$~$ \\
\item[]
\parpic{\includegraphics[width=2cm]{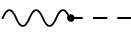}}
$~~\Delta_{\phi_a A_{b\mu}} =0$
$~$ \\
\end{itemize}

We list here the relevant vertices for the one-loop divergent 
ancestor amplitudes (~the factor $\frac{\Lambda^{(D-4)}}{g^2}$ 
is always left understood~)
\begin{itemize}
\item[]
\parpic{\includegraphics[width=2cm]{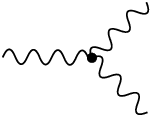}}
$i ~ \G^{(0)}_{A_a^\mu(p_1) A_b^\nu(p_2) A_c^\rho(p_3)} = 
- \epsilon_{abc} \Big [g_{\mu\nu} \big(p_1-p_2 \big)_\rho 
+\\  
~~~~~~~~~~~~~~~~~~~~~~~~~~~\! 
g_{\mu\rho} \big(p_3-p_1\big)_\nu + g_{\nu\rho} \big(p_2-p_3\big)_\mu\Big] $
$~$ \\
\item[]
\parpic{\includegraphics[width=2cm]{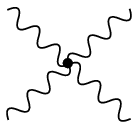}}
$~~ i ~ \G^{(0)}_{A_a^\mu(p_1) A_b^\nu(p_2) A_c^\rho(p_3) A_d^\eta(p_4)} =\\
~~~~~~~~~~~~~~~~
 -i \Big[\, \delta_{ab}\,\delta_{cd}\, 
\big(2 g_{\mu\nu}\,g_{\rho\eta}-g_{\mu\rho}\,g_{\nu\eta}
-g_{\mu\eta}\,g_{\nu\rho}\big)+\\
~~~~~~~~~~~~~~~~~~~~~~~~~ 
\delta_{ac}\,\delta_{bd} \big(-g_{\mu\nu}\,g_{\rho\eta} 
+ 2 g_{\mu\rho}\,g_{\nu\eta}-
g_{\mu\eta}\,g_{\nu\rho}\big) +\\ 
~~~~~~~~~~~~~~~~~~~~~~~~~
\delta_{ad}\,\delta_{bc} \big(- g_{\mu\nu}\,g_{\rho\eta} 
- g_{\mu\rho}\,g_{\nu\eta} +
2 g_{\mu\eta}\,g_{\nu\rho}\big) \Big] $
$~$ \\
\item[]
\parpic{\includegraphics[width=2cm]{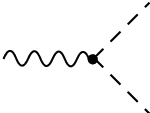}}  
$~$ \\ 
$~~i ~ \G^{(0)}_{A_a^\mu(p_1) \phi_b(p_2) \phi_c(p_3)} = 
\frac{2 M^2}{v^2}\, \epsilon_{abc} \big(p_2 - p_3)_\mu$
$~$ \\ $~$ \\
\item[]
\parpic{\includegraphics[width=2cm]{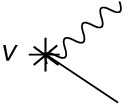}}
$~$ \\
$~~i ~ \G^{(0)}_{A_a^\mu(p_1) V_b^\nu(p_2) B_c(p_3)} = 
-i \, \epsilon_{abc}\, g_{\mu\nu}$
$~$ \\ $~$ \\
\item[]
\parpic{\includegraphics[width=2cm]{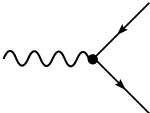}}
$~$ \\
$~~i ~ \G^{(0)}_{A_a^\mu(p_1) c_b(p_2) \bar{c}_c(p_3)} = 
\epsilon_{abc}\, p_{3 \mu}$
$~$ \\ $~$ \\
\item[]
\parpic{\includegraphics[width=2cm]{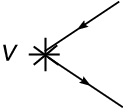}}
$~$ \\
$~~i ~ \G^{(0)}_{V_a^\mu(p_1) c_b(p_2) \bar{c}_c(p_3)} = 
- \epsilon_{abc}\, p_{2 \mu}$
$~$ \\ $~$ \\
\item[]
\parpic{\includegraphics[width=2cm]{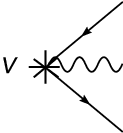}}
$~$ \\
$~~~i ~ \G^{(0)}_{A_a^\mu(p_1) V_b^\nu(p_2) c_c(p_3) \bar{c}_d(p_4)} = 
-i\, g_{\mu\nu}	\, \big( \delta_{ab}\,\delta_{cd} -
\delta_{ad}\, \delta_{bc} \big)$
$~$ \\ $~$ \\
\item[]
\parpic{\includegraphics[width=2cm]{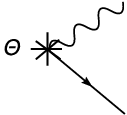}}
$~$ \\
$~~i ~ \G^{(0)}_{A_a^\mu(p_1) \bar{c}_b(p_2) \Theta^\nu_c(p_3)} = 
-i\,\epsilon_{abc}\, g_{\mu\nu}$
$~$ \\ $~$ \\
\item[]
\parpic{\includegraphics[width=2cm]{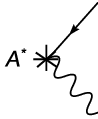}}
$~$ \\
$i ~ \G^{(0)}_{A_a^\mu(p_1) c_b(p_2) A^{* \nu}_c(p_3)} = 
-i\,\epsilon_{abc}\, g_{\mu\nu}$
$~$ \\ $~$ \\ $~$ \\
\item[]
\parpic{\includegraphics[width=2cm]{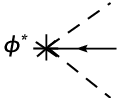}}
$~$ \\ 
$~~i ~ \G^{(0)}_{c_a(p_1) \phi_b(p_2) \phi_c(p_3) \phi^*_d(p_4)} 
= \frac{i}{2v}\,\delta_{ad}\,\delta_{bc}$
$~$ \\ $~$ \\ $~$ \\
\item[]
\parpic{\includegraphics[width=2cm]{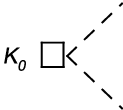}}
$~$ \\
$~~i ~ \G^{(0)}_{\phi_a(p_1) \phi_b(p_2) K_0(p_3)} = 
-\frac{i}{v}\,\delta_{ab}$
$~$\\
\end{itemize}
\begin{itemize}
%
\item $\G^{(1)}[AA]$ 

The relevant graphs are depicted in Figure~\ref{fig.aa}.
\begin{enumerate}
%
\item
\begin{eqnarray}
\frac{9}{2} \frac{1}{(4\pi)^2} \frac{1}{D-4} M^2 \int d^Dx \, A^2 \, .
\label{AA.a}
\end{eqnarray}
%
\item
\begin{eqnarray}
&& 
\!\!\!\!\!\!\!\!\!\!\!\!\!\!\!\!\!\!\!\!
 -6 \frac{1}{(4\pi)^2} \frac{1}{D-4} M^2 \int d^Dx \, A^2 
 - \frac{25}{6} \int d^Dx \, \partial_\mu A_{a\nu} \partial^\mu A^\nu_a
\nonumber \\
&&  
\!\!\!\!\!\!\!\!\!\!\!\!\!\!\!\!\!\!\!\!
+ \frac{14}{3} \cuv \int d^Dx \, \partial A_a \partial A_a \, .
\label{AA.b}
\end{eqnarray}
%
%
\item
\begin{eqnarray}
\frac{1}{12} \cuv \int d^Dx \, \Big (
\partial_\mu A_{a\nu} \partial^\mu A^\nu_a
- \partial A_a^2 \Big ) \, .
\label{AA.c}
\end{eqnarray}
%
%
%
\item
\begin{eqnarray}
-\frac{1}{6} \cuv \int d^Dx \, \Big (
\partial_\mu A_{a\nu} \partial^\mu A^\nu_a
+2  \partial A_a^2 \Big ) \, .
\label{AA.d}
\end{eqnarray}
\end{enumerate}

\begin{figure}
\begin{center}
\includegraphics[width=0.9\textwidth]{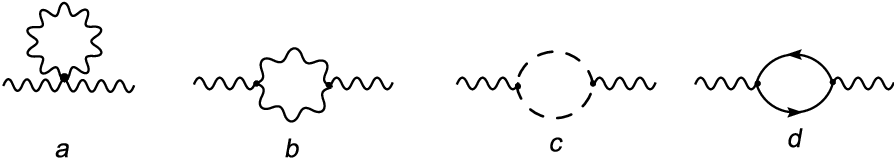}
\end{center}
\caption{Graphs contributing to the 2-point vector meson amplitude}
\label{fig.aa}
\end{figure}

%
\item $\G^{(1)}[AAA]$ 

The relevant graphs are depicted in Figure~\ref{fig.aaa}.
\begin{enumerate}
%
\item
\begin{eqnarray}
- \frac{15}{2} \cuv \int d^Dx \, \epsilon_{abc} \partial_\mu A_{a\nu}
A^\mu_b A^\nu_c \, .
\label{AAA.a}
\end{eqnarray}
%
%
\item 
\begin{eqnarray}
2 \cuv \int d^Dx \, \epsilon_{abc} \partial_\mu A_{a\nu}
A^\mu_b A^\nu_c \, .
\label{AAA.b}
\end{eqnarray}
\item
%
%
\begin{eqnarray}
+ \frac{1}{12} \cuv \int d^Dx \, \epsilon_{abc} \partial_\mu A_{a\nu}
A^\mu_b A^\nu_c \, .
\label{AAA.c}
\end{eqnarray}
\item
%
%
\begin{eqnarray}
- \frac{1}{6} \cuv \int d^Dx \, \epsilon_{abc} \partial_\mu A_{a\nu}
A^\mu_b A^\nu_c \, .
\label{AAA.d}
\end{eqnarray}

\end{enumerate}

\begin{figure}
\begin{center}
\includegraphics[width=0.9\textwidth]{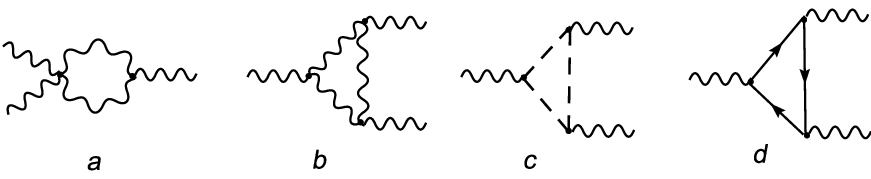}
\end{center}
\caption{Graphs contributing to the 3-point vector meson amplitude}
\label{fig.aaa}
\end{figure}



%
\item $\G^{(1)}[AAAA]$ 

The relevant graphs are depicted in Figure~\ref{fig.aaaa}.
\begin{enumerate}
%
\item
\begin{eqnarray}
- \cuv \int d^Dx 
\, \Big [ \frac{49}{24} (A^2)^2 + \frac{1}{12} (A_{a\mu} A_b^\mu)
(A_{a\nu} A_b^\nu) \Big ]
 \, .
\label{AAAA.a}
\end{eqnarray}
%
%
\item
\begin{eqnarray}
\cuv \int d^Dx \, \Big [ \frac{7}{3} (A^2)^2 + \frac{8}{3} (A_{a\mu} A_b^\mu)
(A_{a\nu} A_b^\nu) \Big ]
 \, .
\label{AAAA.b}
\end{eqnarray}
%
\item
\begin{eqnarray}
- \cuv \int d^Dx \, \Big [ (A^2)^2 + 2 (A_{a\mu} A_b^\mu)
(A_{a\nu} A_b^\nu) \Big ]
 \, .
\label{AAAA.c}
\end{eqnarray}
%
\item
\begin{eqnarray}
- \cuv \int d^Dx \, 
\Big [ \frac{1}{48} (A^2)^2 + \frac{1}{24} (A_{a\mu} A_b^\mu)
(A_{a\nu} A_b^\nu) \Big ]
 \, .
\label{AAAA.d}
\end{eqnarray}
%
%
\item
\begin{eqnarray}
\cuv \int d^Dx \, \Big [ \frac{1}{24} (A^2)^2 + \frac{1}{12} (A_{a\mu} A_b^\mu)
(A_{a\nu} A_b^\nu) \Big ]
 \, .
\label{AAAA.e}
\end{eqnarray}
\end{enumerate}

\begin{figure}
\begin{center}
\includegraphics[width=0.8\textwidth]{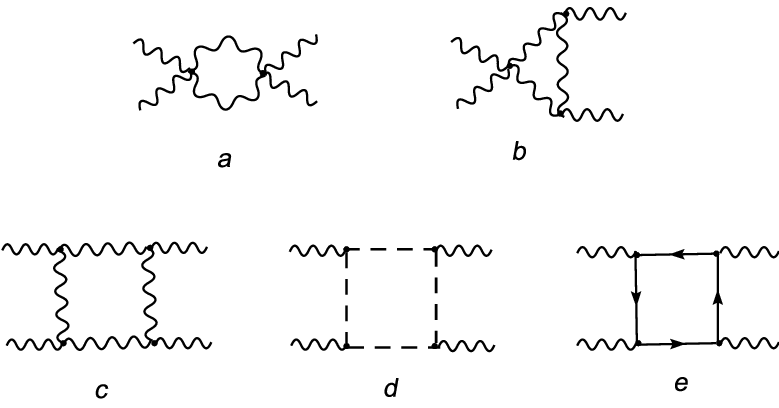}
\end{center}
\caption{Graphs contributing to the 4-point vector meson amplitude}
\label{fig.aaaa}
\end{figure}



%
\item $\G^{(1)}[VA]$ 

The relevant graphs are depicted in Figure~\ref{fig.va}.
\begin{enumerate}
%
\item
\begin{eqnarray}
&& 3 \cuv \int d^Dx \,  M^2 V_{a\mu} A^\mu_a  
+ \frac{8}{3} \cuv \int d^Dx \, V_{a\mu} \square A_a^\mu
\nonumber \\
&& - \frac{5}{3} \cuv \int d^Dx \, V_{a\mu} \partial^\mu \partial A_a 
 \, .
\label{VA.a}
\end{eqnarray}
%
%
\item
\begin{eqnarray}
&& +\frac{1}{3} \cuv \int d^Dx \,\Big [ V_{a\mu} ( \square A^\mu_a 
    - 4 \partial^\mu \partial A_a ) \Big ] \, .
\label{VA.b}
\end{eqnarray}
\end{enumerate}

\begin{figure}
\begin{center}
\includegraphics[width=0.5\textwidth]{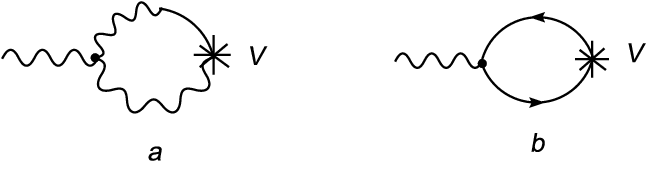}
\end{center}
\caption{Graphs contributing to the 2-point mixed background gauge-vector gauge
 amplitude. The wavy-solid line is the $BA$-propagator}
\label{fig.va}
\end{figure}


%
\item $\G^{(1)}[VAA]$ 

The relevant graphs are depicted in Figure~\ref{fig.vaa}.
\begin{enumerate}
%
\item
\begin{eqnarray}
&& 
\!\!\!\!\!\!\!\!\!\!\!\!\!\!\!\!\!\!
\frac{3}{2} \cuv \int d^Dx \, \Big ( 
\epsilon_{abc}  V_{a\mu} \partial A_b A_c^\mu 
-  \epsilon_{abc}  V_{a\mu} \partial_\nu A_b^\mu A_c^\nu \Big )
 \, .
\label{VAA.a}
\end{eqnarray}
%
%
\item
\begin{eqnarray}
&& \cuv \int d^Dx \,\epsilon_{abc} V_{a\mu} \partial A_b A^\mu_c \, .
\label{VAA.b}
\end{eqnarray}
%
%
\item
\begin{eqnarray}
&& 
\cuv \int d^Dx \,\Big (  
\frac{1}{3} \epsilon_{abc} V_{a\mu} \partial A_b A^\mu_c
- \frac{25}{6} \epsilon_{abc} V_{a\mu} \partial_\nu A_b^\mu A^\nu_c 
\nonumber \\ 
&& ~~~~~~~~~~~~~~~~~~~~~~~
+\frac{17}{6} \epsilon_{abc} V_{a\mu} \partial^\mu A_{b\nu} A^\nu_c
\Big ) \, .
\label{VAA.c}
\end{eqnarray}
%
%

\item
\begin{eqnarray}
&& 
\cuv \int d^Dx \,\Big ( 
\frac{1}{6} \epsilon_{abc} V_{a\mu} \partial A_b A^\mu_c
-\frac{1}{3} \epsilon_{abc} V_{a\mu} \partial_\nu A_b^\mu A^\nu_c 
\nonumber \\ 
&& ~~~~~~~~~~~~~~~~~~~~~~~
+\frac{1}{6} \epsilon_{abc} V_{a\mu} \partial^\mu A_{b\nu} A^\nu_c
\Big ) \, .
\label{VAA.d}
\end{eqnarray}

\end{enumerate}

\begin{figure}
\begin{center}
\includegraphics[width=0.9\textwidth]{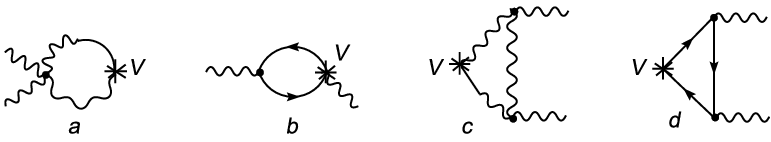}
\end{center}
\caption{Graphs contributing to the one background gauge and two
vector meson legs} 
\label{fig.vaa}
\end{figure}



\item $\G^{(1)}[VAAA]$

The relevant graphs are depicted in Figure~\ref{fig.vaaa}.

\begin{enumerate}
%
\item
\begin{eqnarray} 
- \cuv \int d^Dx \, \Big ( 
 4 V_{a\mu} A^\mu_a A^2 
 -\frac{5}{2} V_{a\mu} A^\mu_b A_{a\nu} A^\nu_b \Big )
 \, .
\label{VAAA.a}
\end{eqnarray}
%
%
\item
\begin{eqnarray}
- \cuv \int d^Dx \, \Big ( 
 \frac{1}{2} V_{a\mu} A^\mu_a A^2 
 +\frac{1}{2} V_{a\mu} A^\mu_b A_{a\nu} A^\nu_b \Big )
 \, .
\label{VAAA.b}
\end{eqnarray}
%
%
\item
\begin{eqnarray}
\cuv \int d^Dx \, \Big ( 
 \frac{4}{3} V_{a\mu} A^\mu_a A^2 
 +\frac{2}{3} V_{a\mu} A^\mu_b A_{a\nu} A^\nu_b \Big )
 \, .
\label{VAAA.c}
\end{eqnarray}
%
%
%
\item
\begin{eqnarray}
\cuv  \int d^Dx \, \Big ( 
 \frac{1}{6} V_{a\mu} A^\mu_a A^2 
 +\frac{1}{3} V_{a\mu} A^\mu_b A_{a\nu} A^\nu_b \Big )
 \, .
\label{VAAA.d}
\end{eqnarray}
\end{enumerate}

\begin{figure}
\begin{center}
\includegraphics[width=0.9\textwidth]{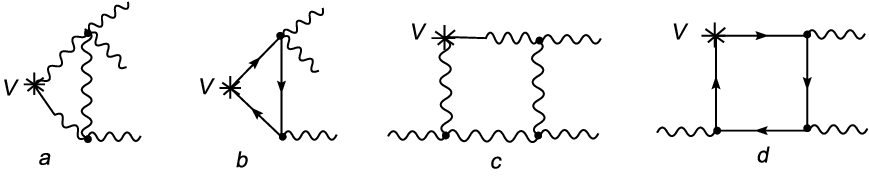}
\end{center}
\caption{Graphs contributing to the one background gauge and three
vector meson legs}
\label{fig.vaaa}
\end{figure}


\item Amplitudes involving an $A^*$-leg

The relevant graphs are depicted in Figure~\ref{fig.atheta}.
\begin{enumerate}
%
\item
\begin{eqnarray} 
-3 \cuv \int d^Dx \, A^*_{a\mu} \Theta^\mu_a
\label{astartheta}
\end{eqnarray}
%
%
\item
\begin{eqnarray}
3\cuv \int d^Dx \, A^*_{a\mu} \partial^\mu c_a
 \, .
\label{astarc}
\end{eqnarray}
%
%
\item
\begin{eqnarray}
\frac{3}{2} \cuv \int d^Dx \epsilon_{abc} A^*_{a\mu} V^\mu_b c_c 
 \, .
\label{astarcv_bolla}
\end{eqnarray}
%
%
%
\item
\begin{eqnarray}
\frac{3}{2} \cuv \int d^Dx \epsilon_{abc} A^*_{a\mu} V^\mu_b c_c 
 \, .
\label{astarc_triang}
\end{eqnarray}
\end{enumerate}

\begin{figure}
\begin{center}
\includegraphics[width=0.9\textwidth]{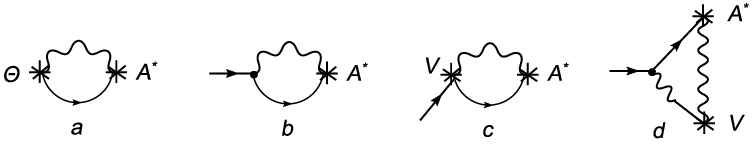}
\end{center}
\caption{Graphs with an $A^*$ leg}
\label{fig.atheta}
\end{figure}


\item Amplitudes involving $K_0, \phi^*$

The relevant graphs are depicted in Figure~\ref{fig.kfstar}.

\begin{enumerate}

\item 
%
%
\begin{eqnarray} 
- \frac{3v^2}{32 M^4} \cuv \int d^Dx \, K_0^2 \, .
\label{k0k0}
\end{eqnarray}
\item 
%
%
%
\begin{eqnarray}
\frac{3v^2}{32 M^4} \cuv \int d^Dx \, K_0 c_a \phi_a^* \, .
\label{k0cphi^*}
\end{eqnarray}
%
%
\item
\begin{eqnarray}
-\frac{3v^2}{128 M^4} \cuv \int d^Dx \, c_a \phi_a^* c_b \phi_b^* \, .
\label{cphistSQ}
\end{eqnarray}
%
%
\item
\begin{eqnarray}
- \frac{v}{8 M^2} \cuv \int d^Dx \, K_0 A^2 \, .
\label{k0aa}
\end{eqnarray}
%
%
\item
\begin{eqnarray}
\frac{v}{16 M^2} \cuv \int d^Dx \, c_a \phi_a^* A^2 \, .
\label{cphistaSq}
\end{eqnarray}
\end{enumerate}

\begin{figure}
\begin{center}
\includegraphics[width=0.8\textwidth]{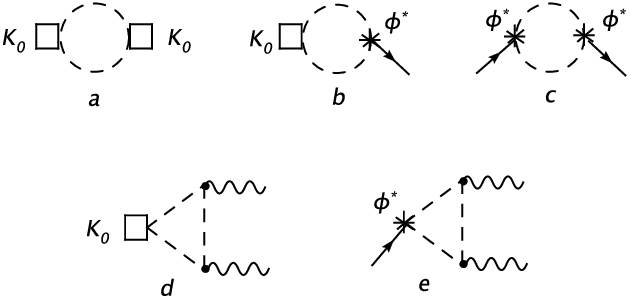}
\end{center}
\caption{Graphs with an $K_0$ and $\phi^*$ legs}
\label{fig.kfstar}
\end{figure}

\end{itemize}
\section{ST identity for the 2-point vector meson amplitude}\label{app:st}

In this Appendix we check the ST identity for
the longitudinal part $\Sigma_L$ of the 2-point vector meson amplitude.

Differentiatiation of the ST identity
\begin{eqnarray}
{\cal S}(\G) & = & \int d^Dx \, \Bigg [ 
\frac{g^2}{\Lambda^{(D-4)}} \Big (
  \frac{\delta \G}{\delta A_{a\mu}^*} \frac{\delta \G}{\delta A_a^\mu}
+\frac{\delta \G}{\delta \phi_a^*}\frac{\delta \G}{\delta \phi_a}
+\frac{\delta \G}{\delta c_a^*} \frac{\delta \G}{\delta c_a} \Big )
\nonumber \\
& &  ~~~~~~~~~ + B_a \frac{\delta \G}{\delta \bar c_a} 
          +\Theta_{a\mu} \frac{\delta \G}{\delta V_{a\mu}} 
          -K_0 \frac{\delta \G}{\delta \phi_0^*} 
 \Bigg ] = 0
\label{st.id}
\end{eqnarray}
w.r.t. $c$, $A_\mu$ yields at one loop
(after setting fields and external sources to zero)
\begin{eqnarray}
&& \G^{(0)}_{c_b(-p) A^*_{c\nu}(p)} 
\G^{(1)}_{A_{c \nu}(-p) A_{a\mu}(p)} +
\G^{(1)}_{c_b(-p) A^*_{c\nu}(p)} ø
\G^{(0)}_{A_{c\nu }(-p) A_{a\mu}(p)} \nonumber \\
&& + \G^{(0)}_{c_b(-p) \phi^*_{c}(p)} 
\G^{(1)}_{\phi_{c}(-p) A_{a\mu}(p)}
+ \G^{(1)}_{c_b(-p) \phi^*_{c}(p)} ø
\G^{(0)}_{\phi_{c}(-p) A_{a\mu}(p)}= 0 \, .
\label{st.id.1}
\end{eqnarray}
By explicit computation one finds
\begin{eqnarray}
&& \G^{(1)}_{\phi_{b}(-p) A_{a\mu}(p)} =0 \, , ~~~~
      \G^{(1)}_{c_b(-p) \phi^*_{a}(p)} = 0 \, , \nonumber \\
&& \G^{(1)}_{c_b(-p) A^*_{a\mu}(p)}  =   \delta_{ab}
p^\mu
\Big [
\frac{1}{2 M^2 p^2} (p^2 + M^2)  H_1(M^2) 
\nonumber \\
&& ~~~~~~~~~~~~~~~~~~~
+ \frac{p^2}{2M^2} H_2(0,0) 
- \frac{(p^2 - M^2)^2}{2 M^2 p^2} H_2(0,M^2) 
\Big ] \, .
\label{st.id.2}
\end{eqnarray}
Moreover
\begin{eqnarray}
&& \G^{(0)}_{A_{b \nu}(-p) A_{a\mu}(p)} =
\delta_{ab}\Big [ \Big (-p^2 + M^2 \Big ) T_{\mu\nu} 
+ M^2 L_{\mu\nu} \Big ] \, , 
\nonumber \\
&& \G^{(0)}_{c_b(-p) A^*_{c\nu}(p)} 
\G^{(1)}_{A_{c \nu}(-p) A_{a\mu}(p)} =
\delta_{ab} ~ i p^\mu \Sigma_L \, . 
\label{st.id.3}
 \end{eqnarray}
By using eqs.(\ref{st.id.2}),
(\ref{st.id.3}) and  the result in eq.(\ref{selfnl.3})
for $\Sigma_L$ one sees that eq.(\ref{st.id.1}) 
is fulfilled.

\end{document}